\documentclass[fleqn,12pt,twoside]{article}
\usepackage{espcrc1}

\usepackage{graphicx}
\usepackage[figuresright]{rotating}

\newcommand{\AmS}{{\protect\the\textfont2
  A\kern-.1667em\lower.5ex\hbox{M}\kern-.125emS}}

\title{Measurement of the neutral pion cross section in proton-proton
	collisions at $\sqrt{s}\!=\!200$~GeV with PHENIX}

\author{H.Torii\address{Kyoto University, Kyoto 606, Japan} for the PHENIX collaboration\thanks{for the full PHENIX Collaboration author list and acknowledgements, see Appendix "Collaborations" of this volume.}}

\begin{document}
\maketitle
\begin{abstract}
The inclusive cross section for neutral pion production
in the range $1\!<\!p_T\!<\!13~{\rm GeV/c}$ in $|\eta|\!<\!0.35$
has been measured by the PHENIX experiment
in proton-proton collisions at $\sqrt{s}\!=\!200~{\rm GeV}$.
An NLO pQCD calculation is, within the experimental and theoretical uncertainties,
consistent with the measurement.
\end{abstract}
\section{Introduction}
A detailed understanding of hadron production in proton-proton
collisions is essential to both the heavy-ion and spin physics
programs at the Relativistic Heavy Ion Collider(RHIC). For the heavy-ion
program, proton-proton data provide the reference to which hadron
production in heavy-ion collisions can be compared so that novel
phenomena, such as jet energy loss or suppression, can be
distinguished from the proton-proton results.
In the spin physics program, hadron
production is a key probe of transverse and longitudinal spin
structure functions and thus an understanding of the unpolarized cross
section with next-to-leading order(NLO) perturbative QCD calculations
provides the theoretical underpinnings for the physics interpretation
of the polarized data.
\section{Experimental Setup}
\begin{figure}[t]
\begin{minipage}[t]{.28\linewidth}
\begin{center}
\includegraphics[width=\linewidth,keepaspectratio]{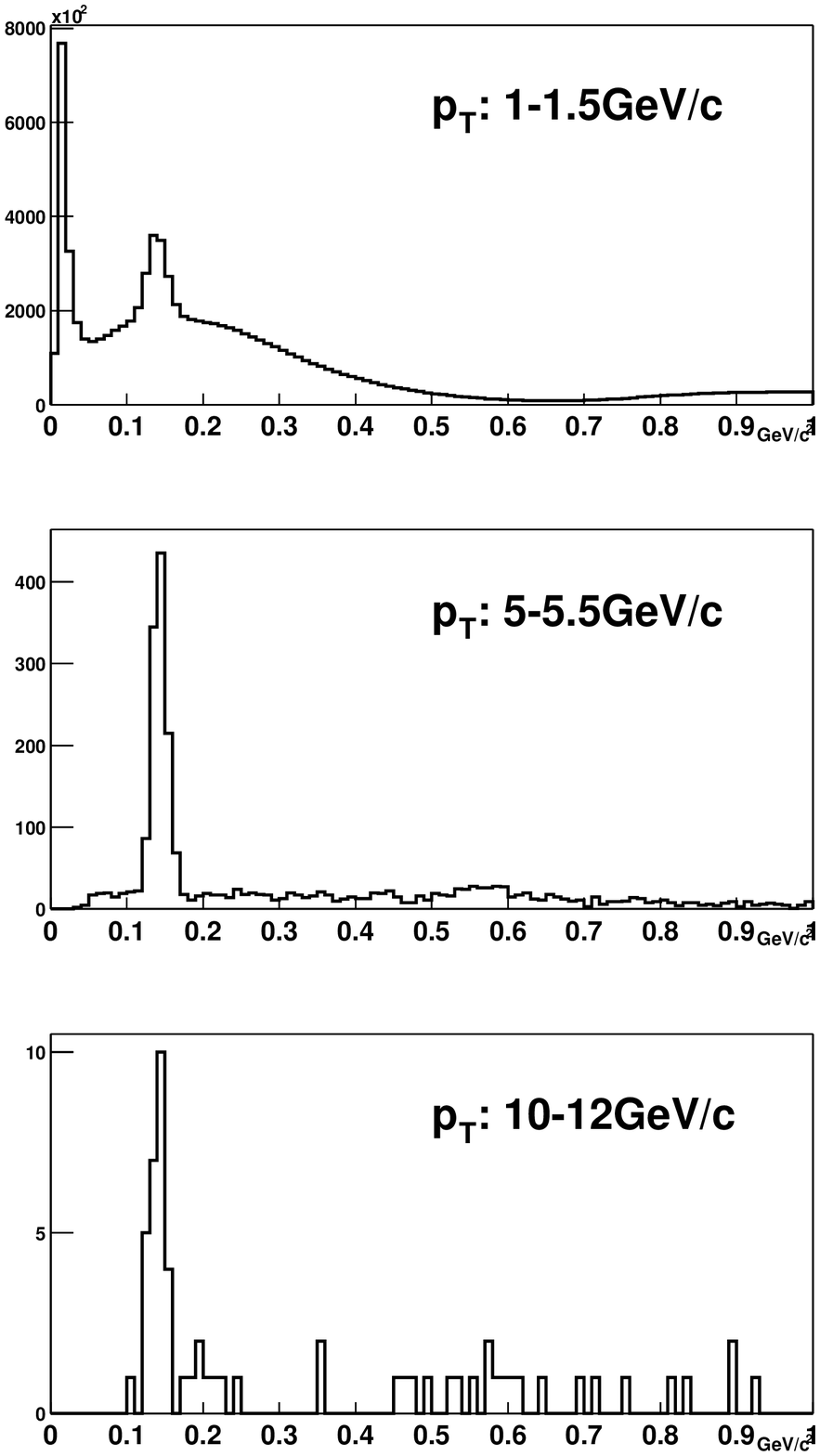}
\end{center}
\vspace{-1cm}
\caption{Invariant mass spectrum for $p_T$ range of
1 to 1.5~GeV/c~(top), 5 to 5.5~GeV/c~(middle), and 10 to 12~GeV/c~(bottom).}
\label{fig:invmass}
\end{minipage}
\hspace{\fill}
\begin{minipage}[t]{.68\linewidth}
\begin{center}
\includegraphics[width=0.95\linewidth,keepaspectratio]{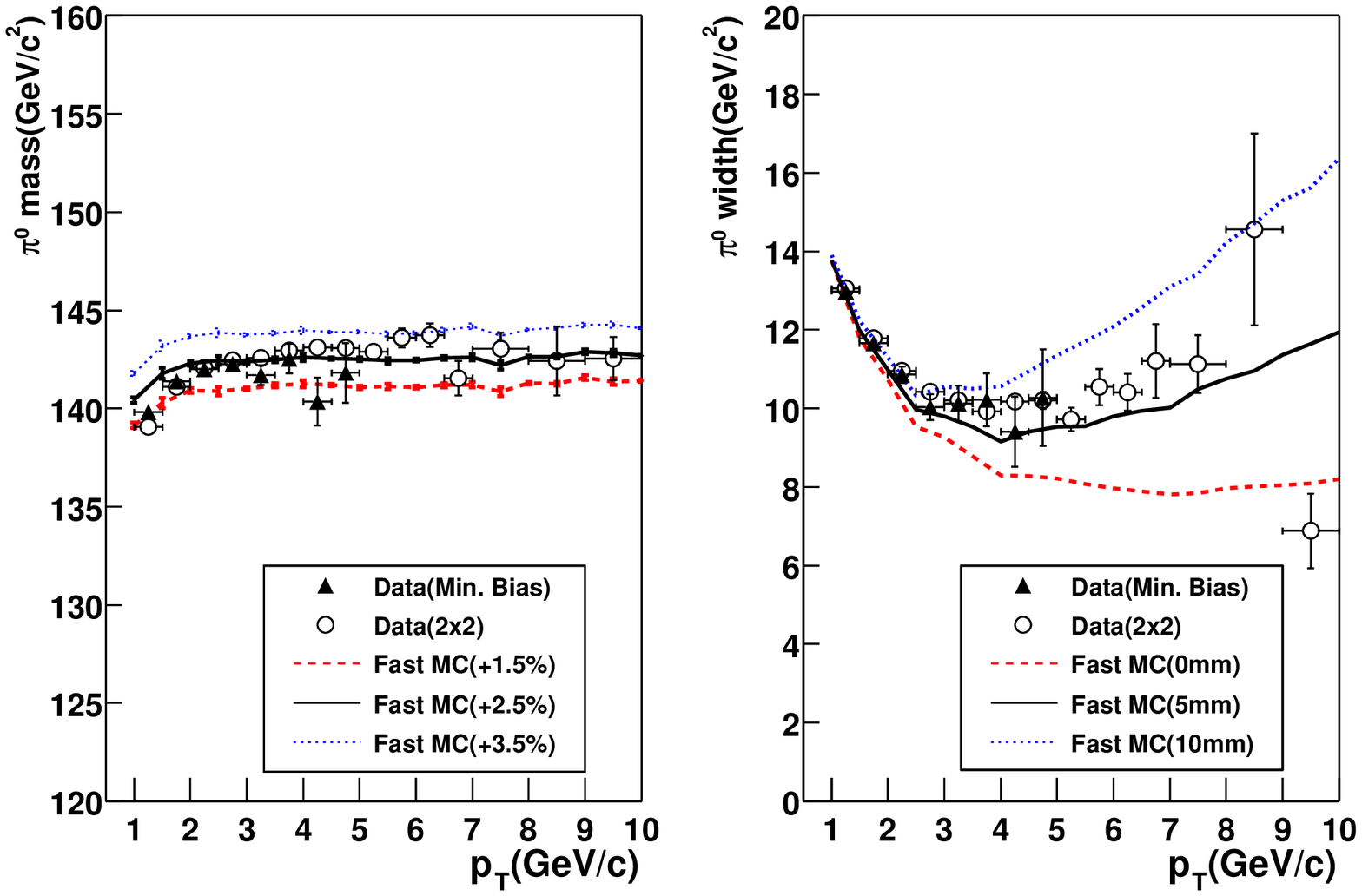}
\vspace{-1.0cm}
\end{center}
\caption{[left] The $\pi^0$ peak position from MB~(star) and 2x2
trigger~(open circle) samples as a function of $p_T$ compared with the Monte Carlo
simulation in which the energy scale is corrected by 1.5\%~(dash), 2.5\%~(solid),
and 3.5\%~(dot), and [right] The $\pi^0$ width compared with the same MC when
the position resolution is increased quadratically from the nominal one by 0~mm~(dash),
5~mm~(solid), and 10~mm~(dot).}
\label{fig:pi0masswidth}
\end{minipage}
\vspace{-0.5cm}
\end{figure}

The PHENIX detector\cite{DM98}\cite{JM02} consists of two central arm 
spectrometers, two muon arm spectrometers, and other detectors
for triggering. This work used the electromagnetic
calorimeters~(EMCal) in the central arms, each
of which have an azimuthal coverage of $90^{\circ}$ and
pseudo-rapidity coverage of $\pm0.35$. 
This detector consists of 6 lead scintillator sampling calorimeter (PbSc)
sectors and 2 lead glass (PbGl) sectors.
In this paper, we will report only the measurement done with the 
5 PbSc sectors,
which have a nominal energy resolution of $8.2\%/\sqrt{E~({\rm GeV})} \oplus 1.9\%$
and a position resolution of $5.7~{\rm mm}/\sqrt{E~({\rm GeV})} \oplus 1.6~{\rm mm}$.

The data were collected during the proton-proton
run in 2001--2002 at RHIC using the
minimum bias~(MB) and the newly installed EMCal triggers.
The MB trigger was made with beam--beam counters~(BBC)
which covered pesudo-rapidity range from 3.0 to 3.9.
The analysis imposed a cut of $\pm30~{\rm cm}$ on the vertex.
The EMCal trigger
was essential to enhance the sample of neutral pions at high
$p_T$. This trigger consisted of two types: 2x2 non-overwrapping
tower sum (0.8~GeV threshold), and 4x4 overwrapping tower sum (2 and
3~GeV thresholds). For higher $p_T$, this work is based upon the data
collected via the 2x2 trigger whose rejection factor was 90.

\section{Analysis Procedure}
The cross sections for the two trigger samples were computed as the ratio of
the ${\pi^0}$ yield corrected for efficiency, acceptance, and smearing
(${\cal N}^{MB}_{\pi^0}$,${\cal N}^{2\times2}_{\pi^0}$) to the
integrated effective luminosity (${\cal L}^{MB}$,${\cal L}^{2\times2}$) 
which were computed as:
\begin{eqnarray}
\hspace{2cm}
{\cal N}^{MB}_{\pi^0} = 
	\frac{N^{MB}_{\pi^0}(p_T) \cdot C^{reco}_{\pi^0}(p_T)}{\varepsilon^{MB}_{\pi^0}(p_T)}
\hspace{2.5cm}
\frac{1}{{\cal L}^{MB}} = \sigma^{pp} \times \frac{\varepsilon^{MB}_{trig}}{N^{MB}_{trig}} \nonumber\\
{\cal N}^{2\times2}_{\pi^0} =
	\frac{N^{2\times2}_{\pi^0}(p_T) \cdot C^{reco}_{\pi^0}(p_T)}
	{\varepsilon^{MB}_{\pi^0}(p_T) \cdot \varepsilon^{2\times2}_{\pi^0}(p_T) }
\hspace{1.3cm}
\frac{1}{{\cal L}^{2\times2}} = \frac{1}{{\cal L}^{MB}} \times
	\frac{N^{MB\&2\times2}_{trig}}{N^{2\times2}_{trig}} \nonumber
\end{eqnarray}

\begin{table}[thb]
\begin{center}
\caption{The $p_{T}$ dependent systematic error.}
\label{tbl:syserr}
\newcommand{\m}{\hphantom{$-$}}
\newcommand{\cc}[1]{\multicolumn{1}{c}{#1}}
\begin{tabular}{|c|c|c|}
\hline
Correction Term		& \m{Source}				& \m{Estimate}   \\
\hline
\hline
$N_{\pi^0}$		& \m{Background subtraction}		& \m{5\%}        \\
			& \m{Hot/Warm towers}			& \m{2-3\%}      \\
			& \m{Run dependence}			& \m{10\%(MB)} \m{6\%(2x2)} \\
\hline
$C^{reco}_{\pi^0}(p_T)$	& \m{Fast MC statistical error}		& \m{1\%}        \\
			& \m{Edge towers}			& \m{5\%}        \\
			& \m{Position resolution}		& \m{0-1\%}      \\
			& \m{Energy absolute calibration}	& \m{3-8\%}      \\
			& \m{Energy non-linearity}		& \m{0-10\%}     \\
			& \m{Energy resolution}			& \m{3\%}        \\
\hline
$\varepsilon^{2\times2}_{\pi^0}(p_T)$	& \m{2x2 high $p_T$ trigger threshold}	& \m{10\%} \\
\hline
\end{tabular}\\[2pt]
\end{center}
\vspace{-1.0cm}
\end{table}

Figure \ref{fig:invmass} shows sample invariant mass spectra for two
photons. Above a $p_T$ of $\sim\!3~{\rm GeV/c}$, the ratio of the
combinatorial background to the signal was $\sim\!10$\%. To extract
the number of $\pi^0$ [$N^{MB}_{\pi^0}(p_T)$ \&
$N^{2\times2}_{\pi^0}(p_T)$] in each $p_T$ bin, several functions --
including a gaussian and some order of polynomial -- were used to fit
to the combinatorial background over a variety of fit ranges. The
systematic error was estimated from the variation in these fits and
the run-to-run stability of the yield.

The acceptance, efficiency, and smearing correction
[$C^{reco}_{\pi^0}(p_T)$] was obtained by a Monte Carlo simulation
which was tuned using results from the test beam measurements and the
data itself. Figure~\ref{fig:pi0masswidth} compares the measured
$\pi^0$ mass and width against the Monte Carlo simulations with
different parameter sets. The systematic error was estimated from the
change in the correction factor when the parameters were varied within
their errors.

Using the MB data, the ${\pi^0}$ efficiency of the 2x2 trigger
[$\varepsilon^{2\times2}_{\pi^0}(p_T)$] was
determined to plateau at 80\% above a $p_T~{\rm of}\sim\!\!3~{\rm GeV/c}$. A systematic error
of 10\% was assigned to this quantity by comparing it to
the result from a Monte Carlo simulation which included
the measured efficiencies for the tiles in the trigger.
The bias for ${\pi^0}$ detection arising from the MB trigger condition
[$\varepsilon^{MB}_{\pi^0}(p_T)$] was measured to be 75\%, independent of $p_T$
up to a $p_T~{\rm of}\sim\!\!5~{\rm GeV/c}$ using the data sample
collected with a 4x4 trigger which, unlike the 2x2 trigger, did not
impose the MB requirement.
This value was consistent with an estimate from a PYTHIA+GEANT simulation and
thus also used to correct the data at higher $p_T$.
All of the systematic errors are summarized in Table~\ref{tbl:syserr}.

The MB trigger efficiency [$\varepsilon^{MB}_{trig}$] of 51\% was obtained
from a PYTHIA+GEANT simulation. We assigned a normalization error of 30\%
based on the difference between the cross section measurement from a 
van der Meer/vernier scan and the total 
(elastic+inelastic) $p\!+\!p$ cross section.
\section{Results and Discussion}
\begin{figure}[htb]
\vspace{-0.7cm}
\begin{minipage}[t]{.49\linewidth}
\includegraphics[width=\linewidth,keepaspectratio]{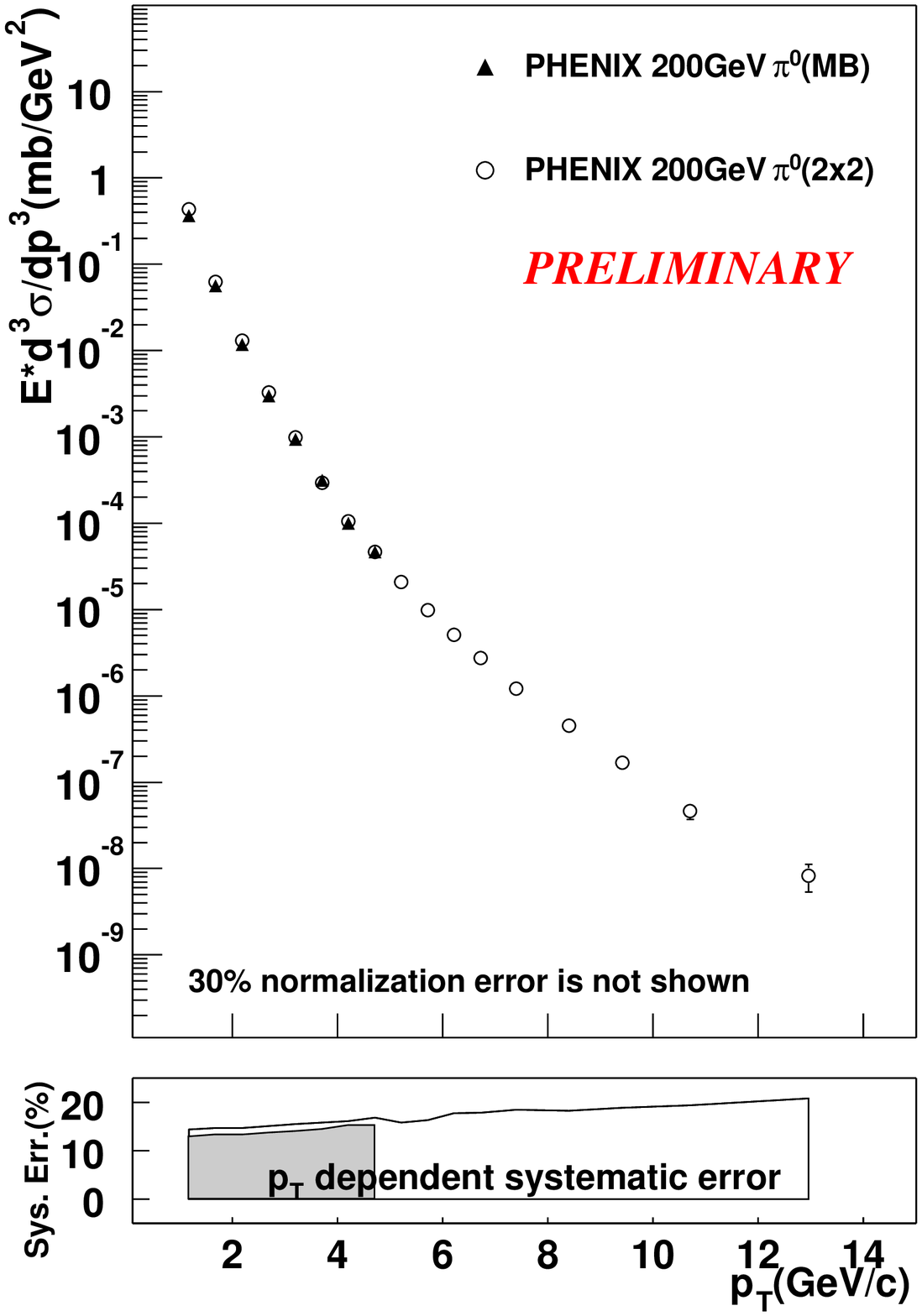}
\vspace{-1.5cm}
\caption{The inclusive neutral pion cross section for the MB
trigger~(filled triangle) and for the 2x2 trigger~(open circle) as a function
of $p_T$. The $p_T$ dependent systematic error is shown in the lower box for
the MB trigger~(filled box) and for the 2x2 trigger~(open box).}
\label{fig:xsect_two}
\end{minipage}
\hspace{\fill}
\begin{minipage}[t]{.49\linewidth}
\includegraphics[width=\linewidth,keepaspectratio]{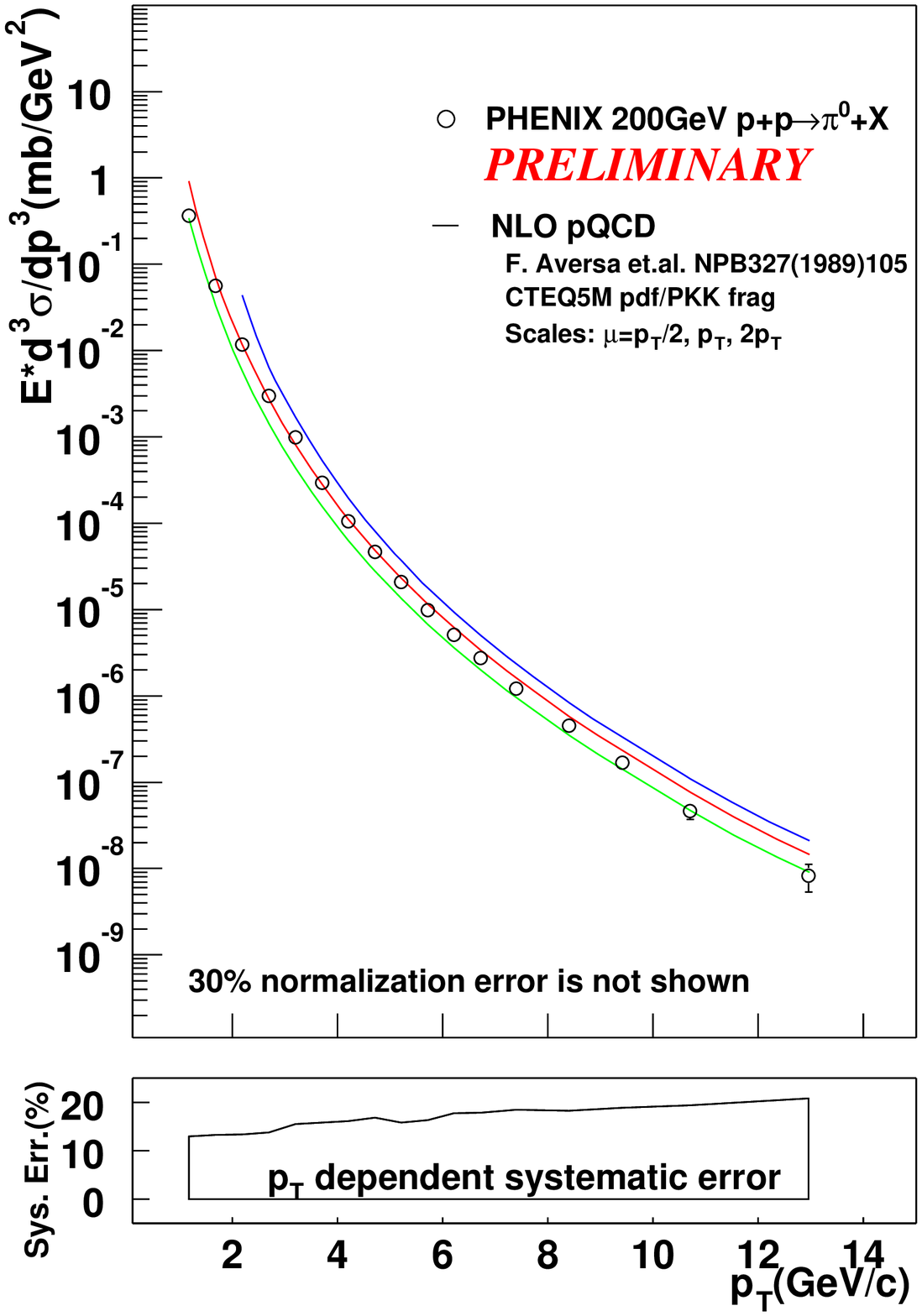}
\vspace{-1.5cm}
\caption{The inclusive neutral pion cross section~(open circle) and
comparison with NLO pQCD calculation using
$p_T/2$~(top line), $p_T$~(middle line), and $2p_T$~(bottom line)
renormalization and factorization scales.
The $p_T$ dependent systematic error of the data is shown in the lower box.}
\label{fig:xsect_theory}
\end{minipage}
\vspace{-0.5cm}
\end{figure}

Figure \ref{fig:xsect_two} shows the measured cross sections for the MB
and the 2x2 trigger samples along with the $p_T$ dependent systematic error.
The results from the two samples are consistent within the error.
The UA1 collaboration measured the $(h^{+}\!+\!h^{-})/2$ production cross section
in the $p\!-\!\bar{p}$ collisions at $\sqrt{s} = 200~{\rm GeV}$ in the $\pm 2.5$
rapidity range\cite{CA80}. When scaled to our rapidity range and corrected for the
particle composition using ISR results\cite{BA75},
our measurement is consistent with the UA1 measurement over their measured $p_T$ range
of 1 to 6~GeV/c.

Figure \ref{fig:xsect_theory} shows
a comparison with an NLO pQCD calculation\cite{WV}
using the formalism of F.~Adversa {\it et al.}\cite{Aver89}
with the CTEQ5M parton distribution functions\cite{CTEQ5M}
and the PKK fragmentation functions\cite{PKK00}.
The data for the lower $p_T$ range is shown
from the MB trigger samples
to avoid the larger systematic error of the 2x2 trigger samples.
Over the full $p_T$ range,
this calculation is consistent with our measurement
within the systematic errors.
\section{Conclusion}
The NLO pQCD calculation with a set of parton distribution and
fragmentation function is consistent with our measurement over
the full $p_T$ range of 1-13~GeV/c within the
systematic error and the scale selection.
This measurement provides a baseline for high $p_T$ heavy-ion physics\cite{DE02}.

\end{document}